\documentclass[b5paper,twoside,reqno]{peloritana}
\usepackage{graphicx}
\usepackage{epstopdf}
\usepackage{amsmath}
\usepackage{amssymb}
\usepackage{lscape}
\usepackage{setspace}
\usepackage{lastpage}
\usepackage{latexsym}
\usepackage{times}
\usepackage{textcomp}
\usepackage[usenames,dvipsnames]{color}
\usepackage{pdfcolmk}
\usepackage{booktabs}
\usepackage[italian,english]{babel}
\usepackage{fancyhdr}
\usepackage{hyperref}

\usepackage[font=small,labelfont={bf}]{caption}
\hypersetup{pdfpagemode={UseOutlines},bookmarksopen,pdfstartview={FitH},colorlinks=true,citecolor=Black,urlcolor=Maroon,linkcolor=black}
\newcommand{\volume}{90}
\newcommand{\issuenumber}{1}
\newcommand{\firstpage}{1}
\newcommand{\publishedyear}{2013}
\newcommand{\doisuffix}{AAPP. ...}
\newcommand{\testata}
{
\vspace*{-5\baselineskip}
\hspace*{-14pt}
{\normalsize{DOI: \href{http://http://dx.doi.org/10.1478/AAPP.91S1A10}{10.1478/\doisuffix}}}\\
\bigskip\\
\hspace*{\fill}{\Large{{AAPP $|$ {Atti della Accademia Peloritana dei Pericolanti}}}\\}
\hspace*{\fill}{\large{{\emph{Classe di Scienze Fisiche, Matematiche e Naturali}}}\\}
\hspace*{\fill}{\href{http://dx.doi.org/10.1478/AAPP.91S1A10 }{\normalsize{ISSN 1825-1242}}\\}
\hspace*{\fill}{\\}
\hspace*{\fill}{{\normalsize{Vol. \volume, Suppl. No. \issuenumber, A10 ~(\publishedyear)\\}}}
}

\makeatletter
\renewcommand{\section}{\@startsection{section}{0}{0mm}{\baselineskip}
{0.5\baselineskip}{\normalfont\normalsize\bfseries}}
\makeatother
\newcommand{\presentazione}{\centerline{\small}}

\input{tcilatex}
\begin{document}

\title{\large Motions in liquid-vapour interfaces\\ by using a continuous mechanical model}
\author{\small\textsc Henri Gouin $^{\star}$}

\begin{abstract}
By using a limit analysis for the  motion equations    of viscous fluid endowed with internal capillarity,  we are able  to propose a dynamical expression for the surface tension of moving liquid-vapour interfaces without any phenomenological assumption. The  proposed relation extends the static case, yields the Laplace formula in cases of mass transfer across interfacial layers and allows to take    the second coefficient of viscosity of compressible fluids into account. We generalize the Maxwell rule in   dynamics and directly explain  the Marangoni effect.
\end{abstract}

\maketitle

\thispagestyle{empty}
\vspace{-0.7cm}
\begin{flushright}
\footnotesize{\em{Dedicated to Professor Giuseppe Grioli on the occasion of his 100th birthday }}
\end{flushright}
%
\section{Introduction}
 Far from the critical point of a fluid,
experimental studies  and spectrography measurements point out that liquid-vapour interfaces have a
thickness of nanometer range and, in
  the vicinity of the layer, vapour and liquid   are homogeneous  \cite{Bongiorno,Ball,Derjaguin,Churaev}.
To model liquid-vapour interfaces, the kinetic theory of gases  proposes fluid equations of state
  as, for example, van der
Waals'  \cite{Rocard,Rowlinson}.  These equations are   correct, more precisely they satisfy the Maxwell rule associated with the isothermal change of phases \cite{Aifantis}. Nonetheless,
they present two main defaults:
\newline
For fluid densities between vapour and liquid, the pressure can be
negative, but simple experiments reveal the existence of  pressures
corresponding to traction in the fluid.
In the domain between vapour and liquid, the internal energy cannot be
represented as a convex surface of the density and entropy; this
fact seems in contradiction with the existence   of two-phase matter
states in stable equilibria \cite{Ono,Meyer}.
\newline
To remove these disadvantages, the thermodynamics usually replaces the non-convex part of the surface energy by a plane domain; but the fluid is not any
more a continuum: the interfacial domain is represented by a material
surface without thickness. Numerous studies related as
well   to fluid mechanics as to thermodynamics interpret  interfaces as
surfaces of discontinuity between two media: a liquid-vapour interface  is usually schematized by a material surface
endowed with a superficial energy. This surface behaves as an
autonomous one \cite{Scriven,Ishi,Defay,Zielinska}. But, this representation is not able to study the
dynamical behavior of the interface more precisely than a surface of
discontinuity and forgets its internal structure.
\newline
In  equilibrium case, it is possible to correct the disadvantages by a
convenient modification of the stress tensor; in the capillary layer  its
expression is anisotropic; the energy of the continuous medium must be
modified.
This \emph{internal capillarity} model is a dynamical theory relevant of second
gradient theory which came from van der Waals and Korteweg \cite{Van der Waals,Korteweg} and was revisited by
Cahn and Hilliard \cite{Cahn}. The model is compatible with the second law of
thermodynamics \cite{Truesdell}. The representation of the internal energy as a function of
the entropy, density and density gradient allows to identify the isothermal
case with a model deduced from molecular theories \cite{Gouin 1,Gouin 2}.
The model is analog to the classical Landau-Ginzburg theory for second order
transition \cite{Landau}.  Far from the critical point the model is qualitative. Nevertheless,
it is a lot more advantageous than the model of Newtonian viscous fluids which is not able to take
 account of layers with a strong gradient of density.
Because we do not consider bubbles and droplets of
radius of some nanometers, the mean surface of interface is of large dimension with respect to its
interfacial thickness \cite{dellIsola}; it is necessary to take account of the different orders of lengths
of our problem: the interfacial layer is of nanometer range, but the radii
of curvature of interfaces are   microscopic.
The surface tension is obtained thanks to the integration across the
capillary layer; it is not necessary uniform along the
interface and depends on the dynamical distributions of density and
temperature.  These distributions take   account of motion equations with a
Navier-Stokes-like   viscosity \cite{Sen}.   At a given temperature, the viscosity
coefficients $\mu $ and   $\eta $ depend on the density ($\mu$ is the dynamic viscosity,  and $\eta $ is the second coefficient  or shear coefficient of viscosity).  For an incompressible fluid, the term involving $\eta $ drops out from the equation; obviously it is not the case through fluid interfaces. We do not assume
any special property on the viscosity coefficients which may strongly vary through
the interface but they are bounded. The dissipative function must have a bounded
integral through the capillary layer and the tangential components of the
velocity field are continuous through the layer \cite{Bedeaux}.
When the temperature distribution is non-uniform
along the interface, the surface tension gradients create a motion along
the capillary layer: this is the so-called Maragoni's effect \cite{Defay}. Thanks to a limit analysis
taking   account of the length ranges of the interface, we are able to
model the Marangoni effect along the interfaces. No special
energy of interface is necessary. When the mass flow across
interface is non-zero, we get a dynamical expression of the Laplace formula.
\\
The proposed method is completely different from the classical calculation
founded on balance equations through a surface of discontinuity  where the
variation of density appears only with a jump through the interface and
when it is necessary to define physical surface quantities as   mass or
entropy per unit of area. Our study is related to interfaces with simple motions.
The calculations are performed in the capillary layer as
in a three-dimensional continuous medium; then, we  consider the limit
case when the interfacial thickness goes towards zero and consequently all the bounded
expressions   have a null integral through the layer \cite{Slemrod}.
We assume that the fluid velocity is bounded together with its partial derivatives
with respect to the coordinates tangent to the interface.
\newline
The model of internal capillarity allows us to obtain a better understanding of
dynamical liquid-vapour interfaces and answer to the question: is the fluid
at the interface rigid or moving \cite{Birkhoff}?
The fluid behavior is different from the classical thermodynamics of
Newtonian fluids: a
supplementary term similar to a heat flux one appears in the equation of energy \cite{Casal1,Dunn}. An integral invariant for motions compatible with the interface consists in
a generalization of the Maxwell rule for isothermal liquid-vapour phase
transition.

In Section 2, we resume the properties of capillary fluids and we develop the
fluid motions in  liquid-vapour interfaces in Sections 3 to 5. A concluding remark focuses on the second coefficient of viscosity.
For the sake of simplicity, all intermediate calculations are proposed in
Appendices as well as some notations.

\section{Equations of motions of viscous fluid endowed with internal
capillarity}

Recall the  main results of a   fluid with internal capillarity \cite{Gouin 1,Casal2}.
We introduce only the specific free energy as a function of the density $%
\rho $, temperature $T$ and $\rm{grad}\,\rho$
\begin{equation*}
\varepsilon =\varepsilon \left( \rho ,T,\beta \right) \text{ \ \ with \ \ }%
\beta =\left( \rm{grad}\,\rho \right) ^{2}.
\end{equation*}%
The specific free energy $\varepsilon $ characterizes together fluid properties
of \emph{compressibility} and \emph{molecular capillarity} of liquid-vapour
interfaces. In accordance with the gas kinetic theory, $\lambda=2\rho \,\varepsilon
_{\beta }^{\prime }(\rho ,\beta )$ is assumed to be constant at a given
temperature  \big ($
\lambda={a\,\kappa^{2}\gamma }/({5\,m^{2}}),
$
where $m$ is the molecular mass of the fluid, $a$\ the internal pressure, $%
\kappa$ the molecular diameter and $\gamma $\ a   factor associated
with molecular potentials of interaction\big) \cite{Rocard,Gouin 2}, and
\begin{equation*}
\rho \,\varepsilon =\rho \,\alpha (\rho )+\frac{\lambda}{2}\,(\rm{grad}\text{\ }%
\rho )^{2},  \label{internal energy}
\end{equation*}%
where the term $({\lambda}/{2})\,(\mathrm{grad\, \rho )^{2}}$ is added to the volume
free energy $\rho \,\alpha (\rho )$ of a compressible fluid.
 Specific free energy
$\alpha $ enables to continuously connect liquid and vapour bulks such that the
pressure $P(\rho )=\rho ^{2}\alpha _{\rho }^{\prime }(\rho )$ is similar to
van der Waals' pressure.
Thanks to experimental data, the $\lambda$ value is    $%
\lambda= 1.17  \times 10^{-5}$ c.g.s.   for water at 20$^{o}$ Celsius \cite{Gouin 3}.
The equation of motion is
\begin{equation}
\rho \ \mathbf{a}=\text{div }\left( \mathbf{\sigma} +{\mathbf{\sigma}}%
_{v}\right) -\rho \;\rm{grad}\text{ }\Omega \ ,  \label{2}
\end{equation}%
where $\mathbf{a}$ is the acceleration vector, $\Omega $ the body force
potential and $\mathbf{\sigma}$ the generalization of the stress tensor:
\begin{equation}
\mathbf{\ \sigma} = -p\,\mathbf{1}-\lambda\;\rm{grad}\text{\, }\rho \ \otimes \
\rm{grad}\, \rho ,  \label{stress tensor}
\end{equation}%
with $p=\rho ^{2}\varepsilon _{\rho }^{\prime }-\rho \text{ div\textrm{\ }}(\lambda%
\text{ grad }\rho )$; the viscous stress tensor is \cite{Landau}
\begin{equation*}
\mathbf{\sigma }_{v}=\eta\,  (\text{tr\,}{\textbf{D}}) \,\mathbf{1}+2\mu \;{\textbf{D}},
\end{equation*}%
where ${\textbf{D}}$ denotes the velocity strain tensor.
Equation   (\ref{2}) can be written in the form
\begin{equation}
\rho \ \mathbf{a} +{\rm grad}\,\emph{P}+ \rho\, { \rm grad}\,\omega -{\rm
div }{\mathbf{\sigma }}_{v}\ =0,  \label{3}
\end{equation}%
where $\omega =\Omega -\lambda\,\Delta \rho $. The equation of motion must be
completed by the balance of mass
\begin{equation}
\frac{\partial \rho }{\partial t}+\rm{div}\left( \rho\, \mathbf{u}\right) =0,
\label{4}
\end{equation}
where $\mathbf{u}$ is the velocity vector.
Let us note that the equation of energy can be written in the form \cite{Casal1,Dunn}
\begin{equation*}
\frac{\partial e}{\partial t}+{\rm div}\left[ \left( e\,\mathbf{1} -\mathbf{\sigma }-%
\mathbf{\sigma }_{v}\right) \mathbf{u}\right] \mathbf{-\rm{div}}\left( \lambda\,
\frac{d\rho }{dt}\,\rm{grad}\rho \right) +{\rm{div}}\,  \textbf{q} - \emph{r}-\rho \frac{\partial
\Omega }{\partial t}=0,
\end{equation*}%
where $e = \rho\left(\frac{1}{2}\textbf{u}^2 + \varepsilon + \Omega\right)$, $\textbf{q}$ is the heat flux vector and $r$ the heat supply,
such that the model  is compatible with the
second law of thermodynamics \cite{Truesdell}.

\section{The dynamical surface tension}
\noindent \emph{Now, for the sake of simplicity, we neglect the body forces.}
\subsection{Case of a planar interface at equilibrium}
The eigenvalues of the stress tensor in internal capillarity are deduced
from Eq. (\ref{stress tensor}):
\newline
$\lambda _{1}=-p+\lambda\left( \rm{grad}\rho \right) ^{2}$ is the eigenvalue
associated with the plane orthogonal to $\rm{grad}\,\rho $,
\newline
$\lambda _{2}=-p$ is the eigenvalue associated with the direction of $\rm{%
grad}\,\rho $.
\newline
The classical notations are presented in Appendix 1; in the system of coordinates associated with the interface, the stress
tensor can be written
\begin{equation*}
\mathbf{\sigma} = \left[
\begin{array}{ccc}
\lambda _{1} & 0 & 0 \\
0 & \lambda _{1} & 0 \\
0 & 0 & \lambda _{2}%
\end{array}%
\right] .
\end{equation*}
The equation of equilibrium of the planar interface is deduced from Eq. (\ref%
{2}) and by neglecting the body forces, we get
\begin{equation*}
\lambda _{2}=-P_{o},
\end{equation*}
where $P_{o}$ denotes the common pressure in the vapour and liquid bulks.
Per unit of length, the line force exerted on the edge of the interface is
\begin{equation*}
F=\int_{x_{_3}^{v}}^{x_{_3}^{l}}\lambda _{1}h_{_3}\
dx_{_3}=-P_{o}\,h+\int_{x_{_3}^{v}}^{x_{_3}^{l}}\lambda\left( \rm{grad}\,\rho \right)
^{2}h_{_3}\ dx_{_3},
\end{equation*}
where the subscript $3$
denotes the normal component to the density surfaces    of the capillary
layer, $h$ denotes the interface thickness and $l$, $v$ indicate the liquid
and vapour bulks. In the limit analysis of thin interfaces, the term $P_{o}\,h$ is negligible. Let us denotes
$\displaystyle
H=\int_{x_{_3}^{v}}^{x_{_3}^{l}}\lambda\left( \rm{grad}\,\rho \right) ^{2}h_{_3}
dx_{_3}.
$
The line force $H$ exerted per unit of length  corresponds to the
surface tension.

\subsection{The dynamical surface tension value}

The notations are presented in Appendices 1 and 2.
The equation of motion (\ref{3}) is separated into normal and
tangential components. In the orthogonal coordinate system presented in
appendix 1,
\begin{eqnarray}
\rho \, a _{tg}+{\rm{grad}_{tg}}\,P &=&\rho\, \lambda\,{\rm{grad}_{tg}}\,\Delta \rho +%
{\rm{grad}_{tg}}\left(\eta\, \rm{div}\,\mathbf{u}\right) +2\,\rm{div}
\left( \mu\, \textbf{D}\right) _{tg}  \label{10}, \\
\rho \, a_{_3}+\frac{1}{h_{_3}}\frac{\partial P}{\partial x_{_3}} &=&\rho\, \lambda%
\frac{1}{h_{_3}}\frac{\partial \Delta \rho }{\partial x_{_3}}+\frac{1}{h_{_3}}%
\frac{\partial \left( \eta\, \rm{div}\mathbf{u}\right) }{\partial x_{_3}}+2\,
\rm{div}\left( \mu\, \textbf{D}\right) _{3}  \label{11},
\end{eqnarray}%
where the subscript $tg$ denotes the tangential component  to the density surfaces    of the capillary
layer. The normal vector $\mathbf{e}_{_3}$ corresponds to the direction of the increasing
densities. An integration of Eq. (\ref{11}) across the interface yields
\begin{eqnarray*}
\int_{x_{_3}^{v}}^{x_{_3}}\rho \  {a}_{_3}\,h_{_3}
dx_{_3}+\int_{x_{_3}^{v}}^{x_{_3}}\frac{\partial P}{\partial x_{_3}}\,
dx_{_3}=\int_{x_{_3}^{v}}^{x_{_3}}\rho\, \lambda\, \frac{\partial \Delta \rho }{\partial
x_{_3}}\,dx_{_3}\\ +\int_{x_{_3}^{v}}^{x_{_3}}\frac{\partial \left( \eta\, \rm{div}%
\mathbf{u}\right) }{\partial x_{_3}}\,dx_{_3}+2\int_{x_{_3}^{v}}^{x_{_3}}\rm{div}%
\left( \mu\, \textbf{D}\right) _{3}h_{_3}  dx_{_3}.
\end{eqnarray*}
The fluid is assumed to cross   the capillary layer.  Taking
account of Eq. (\ref{A14}) and Eq. (\ref{A23}) proved in Appendix 2, we get,
\begin{equation*}
P-P_{v}+Q^{2}\left(\frac{1}{\rho}-\frac{1}{\rho _{v}}\right)=\lambda \rho\, \Delta
\rho\,-\lambda \int_{x_{_3}^{v}}^{x_{_3}}\Delta \rho  \frac{\partial \rho }{\partial
x_{_3}} dx_{_3}+\left[ \left(\eta+2 \mu \right)D_{_{33}}\right]
_{x_{_3}^{v}}^{x_{_3}}-\frac{2}{R_{m}}\left[ \eta\, u_{_3}\right]
_{x_{_3}^{v}}^{x_{_3}},
\end{equation*}
where $Q$ is the mass flow across the capillary layer,  $R_{m}$ the
mean radius of curvature of the surfaces of equal density oriented following
$\mathbf{e}_{_3}$ and $[ \ ]$ denotes  the difference of values through the interface. Taking  account of Eqs. (\ref{A4},\ref%
{A8},\ref{A9},\ref{A10}), we obtain
\begin{eqnarray*}
P-P_{v}+Q^{2}\left( \frac{1}{\rho }-\frac{1}{\rho _{v}}\right)  =\lambda\,\rho\,
\Delta \rho +\lambda\int_{x_{_3}^{v}}^{x_{_3}}\frac{2}{R_{m}}\frac{1}{h_{_3}^{2}}%
\left( \frac{\partial \rho }{\partial x_{_3}}\right) ^{2}h_{_3}  dx_{_3}\\
-\frac{\lambda
}{2}\left[ \frac{1}{h_{_3}^{2}}\left( \frac{\partial \rho }{\partial x_{_3}}%
\right) ^{2}\right] _{x_{_3}^{v}}^{x_{_3}} -Q\left[ \left( \eta +2\mu \right) \frac{1}{h_{_3}}\frac{1}{\rho ^{2}}%
\frac{\partial \rho }{\partial x_{_3}}\right] _{x_{_3}^{v}}^{x_{_3}}-\frac{2}{%
R_{m}}\left[ \eta u_{_3}\right] _{x_{3}^{v}}^{x_{_3}},
\end{eqnarray*}
where $u_{_3}$ is the third component of $\textbf{u}$.
The radius of curvature $R_{m}$ is assumed to be constant across the
capillary layer (see Appendix 1); then,
\begin{eqnarray}
P-P_{v}+Q^{2}\left( \frac{1}{\rho }-\frac{1}{\rho _{v}}\right) &=&\lambda\,\left\{
\rho\, \Delta \rho -\frac{1}{2}\left( \rm{grad}\,\rho \right) ^{2}\right\} +%
\frac{2\,\lambda}{R_{m}}\int_{x_{_3}^{v}}^{x_{_3}}\left( \rm{grad}\,\rho
\right) ^{2}h_{_3}  dx_{_3}  \notag \\
&&-Q\left\{ \left[ \left( \eta +2\,\mu \right) \frac{\mathbf{e}_{_3}.\rm{%
grad}\rho }{\rho ^{2}}\right] _{x_{_3}^{v}}^{x_{_3}}+\frac{2}{R_{m}}\left[
\frac{\eta }{\rho }\right] _{x_{_3}^{v}}^{x_{_3}}\right\} . \label{12}
\end{eqnarray}%
The terms $\lambda\left\{ \rho\, \Delta \rho -\frac{1}{2}\left( \rm{grad}\rho
\right) ^{2}\right\} $ and $Q\left\{ \left[ \left( \eta +2\mu \right)
\left({\mathbf{e}_{_3}.\rm{grad}\,\rho}\right)/{\rho ^{2}}\right]
_{x_{_3}^{v}}^{x_{_3}^{l}}\right\} $ are null in the liquid and vapour bulks.
Consequently, we get:
\begin{equation}
P_{l}-P_{v}=Q^{2}\left( \frac{1}{\rho _{v}}-\frac{1}{\rho _{l}}\right) +%
\frac{2\ K}{R_{m}},   \label{13}
\end{equation}%
where%
\begin{equation}
K=H-Q\left( \frac{\eta_{l}}{\rho _{l}}-\frac{\eta_{v}}{\rho
_{v}}\right)\qquad {\rm with}\qquad H=\lambda\int_{x_{_3}^{v}}^{x_{_3}}\left( \rm{grad}\,\rho \right) ^{2}h_{_3}\ dx_{_3} .\label{14}
\end{equation}%
Equation (\ref{13}) extends the Laplace formula which is obtained when $Q=0$.
The term $H$ can be interpreted as the \emph{dynamical surface tension} of
an interface crossed by a viscous fluid and $K$ as the \emph{viscous
dynamical surface tension}. The surface tension depends on the dynamical
distribution of the density through the interface
and on the volume viscosity $\eta$, only. For a plane interface,
\begin{equation}
P_{l}-P_{v}=Q^{2}\left( \frac{1}{\rho _{v}}-\frac{1}{\rho _{l}}\right)
\label{16}.
\end{equation}%
Equation (\ref{16}) expresses the equality of normal stresses on an
interface crossed by viscous fluid and classically obtained in the
literature. In the case when $H=0$, Relations  (\ref{13})   and (\ref{16}) cannot
be identified with shock conditions. In a dissipative flow with a
domain with strong gradients of density schematized in perfect fluid by a shock
wave, the fluid is weakly dissipative and the relations of discontinuity are
expressed in form of expansion with respect to the inverse of the Reynolds
number \cite{Germain}.
\section{Practical calculus of the surface tension}

Let us consider the case when \emph{the flux of mass is null across the interface}.
The capillary layer is subject to tangential motions.\newline \newline
\textbf{Definition}
: \emph{A motion is compatible with the capillary layer if the surfaces of density
are material surfaces.}
\newline
In the capillary layer\
$
d\rho /{dt}=0
$\ and
consequently,
$
\rm{div}\mathbf{\,u}=0.
$
  Then, Eq. (\ref{3})
can be written
\begin{equation*}
\rho \ \mathbf{a} +{\rm{grad}}\,P=\lambda\,\rho\,  {\rm{grad}}\,\Delta \rho +%
 2 \,\text{div}\left( \mu\, \textbf{D}\right).
\end{equation*}%
Equation (\ref{11}) yields
\begin{equation*}
a_{_3}+\frac{1}{\rho h_{_3}}\,\frac{\partial P}{\partial x_{_3}}=\frac{\lambda}{%
h_{_3}}\frac{\partial \Delta \rho }{\partial x_{_3}}+\frac{2}{\rho }\,\rm{div}%
\left( \mu\, \textbf{D}\right) _{3},
\end{equation*}
and we get
\begin{equation*}
\int_{x_{3}^{v}}^{x_{3}}a_{_3}h_{_3}dx_{3}+
\int_{x_{_3}^{v}}^{x_{3}}\frac{1}{\rho}\, \frac{\partial P}{\partial x_{_3}}\,
dx_{_3}=\int_{x_{_3}^{v}}^{x_{_3}}\lambda\,\frac{\partial \Delta \rho }{\partial x_{_3}}\,
dx_{3}+\int_{x_{_3}^{v}}^{x_{_3}}\frac{2}{\rho }\  {\rm{div}}\left( \mu\, \textbf{D}\right)
_{3}h_{3}  dx_{3}.
\end{equation*}%
For the limit analysis of thin interfaces, two terms are null and Eqs. (\ref{A21}, \ref{A25}) in Appendix 2 yield
\begin{equation}
\lambda\,\Delta \rho =\frac{P}{\rho}-\frac{P_{v}}{\rho_{v}}+\int_{\rho _{v}}^{\rho
}\frac{P}{\rho ^{2}}\,d\rho .  \label{18}
\end{equation}%
Consequently, in the liquid bulk,%
\begin{equation}
\int_{\rho _{v}}^{\rho _{l}}\frac{P}{\rho ^{2}}\,d\rho =\frac{P_{v}}{\rho _{v}}%
-\frac{P_{l}}{\rho _{l}} .\label{19}
\end{equation}
Relation (\ref{19}) is an integral invariant associated with motions
compatible with the capillary layer. In the special case of isothermal
equilibrium, we get Eq. (4-11) from \cite{Aifantis}. In the plane case, we are back
to the Maxwell rule of equality of areas.
Equation  (\ref{18}) writes
\begin{equation}
\lambda\,\Delta \rho =\frac{\partial}{\partial \rho }\left( \rho\, \int_{\rho
_{v}}^{\rho}\frac{P-P_{v}}{\rho ^{2}}\,d\rho \right).  \label{20}
\end{equation}%
To a constant temperature, $\rho \int_{\rho _{v}}^{\rho _{l}}\left(
P-P_{v}\right) / \rho ^{2} d\rho $ is the Helmoltz free energy per unit
volume of the fluid.
If we assume a regular variation of the temperature in the capillary layer, $%
\left( \partial P/\partial \theta \right) \left( \partial \theta /\partial
x_{_3}\right) $ is negligible with respect to $\left( \partial P/\partial
\rho \right) \left( \partial \rho /\partial x_{_3}\right) $.\ By taking
account of Eq. (\ref{A4}), Eq. (\ref{20}) yields
\begin{equation*}
-\frac{2}{R_{m}}\frac{\lambda}{h_{_3}}\left( \frac{\partial \rho }{\partial x_{_3}}%
\right) ^{2}+\lambda\,\left( \frac{1}{h_{_3}}\frac{\partial \rho }{\partial x_{_3}}%
\right) \left( \frac{1}{h_{_3}}\frac{\partial \rho }{\partial x_{_3}}\right)
_{,3}=\frac{\partial }{\partial x_{_3}}\left( \rho \int_{\rho _{v}}^{\rho }%
\frac{P-P_{v}}{\rho ^{2}}\,d\rho \right).
\end{equation*}%
An integration across the capillary layer yields
\begin{equation}
\frac{\lambda}{2}\left(\rm{grad}\rho \right) ^{2}=\frac{2\, \lambda}{R_{m}}%
\int_{x_{_3}^{v}}^{x_{_3}}\left(\rm{grad}\rho \right)^{2}h_{_3}
dx_{_3}+\rho \int_{\rho _{v}}^{\rho }\frac{P-P_{v}}{\rho ^{2}}\,d\rho.
\label{21}
\end{equation}
All the same,
\begin{equation*}
\frac{\lambda}{2}\left( \rm{grad}\rho \right) ^{2}=\frac{2\, \lambda}{R_{m}}%
\int_{x_{_3}^{l}}^{x_{_3}}\left(\rm{grad}\rho \right) ^{2}h_{_3}
dx_{3}+\rho \int_{\rho _{l}}^{\rho }\frac{P-P_{v}}{\rho ^{2}}\,d\rho.
\label{22}
\end{equation*}
Let us denote by $x_{_3}^{i}$ the third coordinate of a surface of density $\rho _{i},\  (\rho _{i}\equiv\frac{1}{2}\left( \rho _{v}+\rho _{l}\right)) $. Due to the fact we assume the radius of curvature   of non-molecular
size, for   {$x_{_3}\in\left[ x_{_3}^{v},x_{_3}^{i}\right]$}  the quantity $%
\left( 2\, \lambda/R_{m}\right) \int_{x_{_3}^{v}}^{x_{_3}}\left(\rm{grad}\rho
\right) ^{2}h_{_3}  dx_{3}$ is negligible with respect to $({\lambda}/{2})\,\left(\rm{grad}\rho \right) ^{2}$.  Then,
\begin{equation}
\text{for}\ \rho \in \left[ \rho _{v},\rho _{i}\right] ,\ \ \frac{\lambda}{2}%
\left( \rm{grad}\rho \right) ^{2}=\rho \int_{\rho _{v}}^{\rho }\frac{%
P-P_{v}}{\rho ^{2}}\,d\rho.  \label{23}
\end{equation}
All the same,
\begin{equation}
\text{for}\ \rho \in \left[ \rho _{i},\rho _{l}\right] ,\ \ \frac{\lambda}{2}%
\left( \rm{grad}\rho \right) ^{2}=\rho \int_{\rho _{l}}^{\rho }\frac{%
P-P_{v}}{\rho ^{2}}\,d\rho.  \label{24}
\end{equation}
Relations (\ref{14}), (\ref{23}) and (\ref{24}) yield
\begin{equation*}
H=2\left\{ \int_{x_{_3}^{v}}^{x_{_3}^{i}}\rho \left( \int_{\rho _{v}}^{\rho }%
\frac{P-P_{v}}{\rho ^{2}}\,d\rho \right) h_{_3}\,
dx_{_3}+\int_{x_{_3}^{i}}^{x_{_3}^{l}}\rho \left( \int_{\rho _{l}}^{\rho }\frac{%
P-P_{v}}{\rho ^{2}}\,d\rho \right) h_{_3}\, dx_{_3}\right\}.
\end{equation*}
Taking  $d\rho =h_{_3}\sqrt{\left( \rm{grad}\rho \right)
^{2}}\,dx_{_3}$ into account, we get
\begin{equation}
H=\sqrt{2\,\lambda}\left\{ \int_{\rho _{v}}^{\rho _{i}}\left( u\sqrt{\int_{\rho
_{v}}^{u}\frac{P-P_{v}}{\rho ^{2}}\,d\rho }\right) \, du+\int_{\rho _{i}}^{\rho
_{l}}\left( u\sqrt{\int_{\rho _{l}}^{u}\frac{P-P_{l}}{\rho ^{2}}\,d\rho }%
\right) \, du\right\}.  \label{25}
\end{equation}
Expression (\ref{25}) allows to calculate the surface tension of a moving
capillary layer; pressure $P$ is a function of $\rho $ and $\theta $ in
each point of the layer.
\newline
Let us note that for the limit case when the capillary layer thickness is
null, the viscosity of the fluid does not explicitly appear in Rel. (\ref
{25}).  In the isothermal case of a planar interface at equilibrium, Rel. (
\ref{25}) is equivalent to
\begin{equation*}
H=\sqrt{2\,\lambda}\int_{\rho _{v}}^{\rho _{l}}\sqrt{f(\rho )}\,d\rho,
\end{equation*}
where $f(\rho )$ is the free energy per unit volume which is null for $\rho
=\rho _{v}$ and $P_{v}=P_{l}=P_{o}$.  The $H$ value of is  numerically calculable
by using
thermodynamical pressure models through interfaces in the form  $P=P(\rho ,\theta
)$.
\section{Marangoni effect for liquid-vapour interfaces}

The conditions of our study are the same than in Section 4: \emph{the flux of mass
across the interface is null; the surfaces of density are material surfaces}.
Equation (\ref{12}) yields
\begin{equation*}
P-P_{v}=\lambda\left\{ \rho\, \Delta \rho -\frac{1}{2}\left( \rm{grad}\rho \right)
^{2}\right\} + \frac{2}{R_{m}}\,H_{v}\left( x_{_3}\right)  , \label{27}
\end{equation*}%
where%
\begin{equation*}
H_{v}\left( x_{_3}\right) =\lambda\,\int_{x_{_3}^{v}}^{x_{_3}}\left(\rm{grad}\rho
\right) ^{2}h_{_3} dx_{3}.
\end{equation*}%
All the same,
\begin{equation*}
P-P_{l}=\lambda\left\{ \rho\, \Delta \rho -\frac{1}{2}\left( \rm{grad}\rho \right)
^{2}\right\} +\frac{2}{R_{m}}\,H_{l}\left( x_{_3}\right),   \label{28}
\end{equation*}%
where
\begin{equation*}
H_{l}\left( x_{_3}\right) =\lambda\int_{x_{_3}}^{x_{_3}^{l}}\left( \rm{grad}\rho
\right) ^{2}h_{_3}  dx_{_3}.
\end{equation*}%
If we transfer these results into Eq. (\ref{10}),   we obtain for $j\in \left\{
v,l\right\} $,
\begin{equation}
\rho \, a_{tg}=\frac{\lambda}{2}\,{\rm{grad}_{tg}}\left( \rm{grad}\rho
\right) ^{2}-{\rm{grad}_{tg}}\left(\frac{2}{R_{m}}\,H_{j}\left( x_{_3}\right)\right)
+{\rm{grad}_{tg}}\left( P_{j}\right) +2\,\rm{div}\left( \mu\, \textbf{D}\right) _{tg}.
\label{29}
\end{equation}%
By integration of Eq. (\ref{29})\ across the capillary layer,
\begin{eqnarray}
\int_{x_{_3}^{v}}^{x_{_3}^{l}}\rho \, a_{tg}h_{_3}   dx_{_3}
&=&\int_{x_{_3}^{v}}^{x_{_3}^{l}}\frac{\lambda}{2}\,{\rm{grad}_{tg}}\left(\rm{grad}%
\rho \right) ^{2}h_{_3}  dx_{_3}-\int_{x_{_3}^{v}}^{x_{_3}^{i}}{\rm{grad}}
_{tg}\left( \frac{2}{R_{m}}\,H_{v}\left( x_{_3}\right) \right) h_{_3} dx_{_3}  \notag\\
&&-\int_{x_{_3}^{v}}^{x_{_3}^{i}}{\rm{grad}_{tg}}\left(P_{v}\right) h_{_3}
dx_{_3}-\int_{x_{_3}^{i}}^{x_{_3}^{l}}{\rm{grad}_{tg}}\left( \frac{2}{R_{m }}H_{l}\left( x_{_3}\right) \right) h_{_3}  dx_{_3}\notag  \\
&&-\int_{x_{_3}^{i}}^{x_{_3}^{l}}{\rm{grad}_{tg}}\left( P_{l}\right) h_{_3}
dx_{_3}+2\int_{x_{_3}^{v}}^{x_{_3}^{l}}\rm{div}\left( \mu\, \textbf{D}\right)
_{tg}h_{_3}  dx_{_3}.  \label{30}
\end{eqnarray}%
Equation (\ref{21}) yields
\begin{equation*}
\frac{\lambda}{2}\left(\rm{grad}\,\rho \right) ^{2}=\frac{2}{R_{m}}\,
H_{v}\left( x_{_3}\right) +\rho  \int_{\rho _{v}}^{\rho }\frac{P-P_{v}}{%
\rho ^{2}}\,d\rho  ,
\end{equation*}%
and by integration,
\begin{equation*}
H_{v}\left( x_{_3}\right) =\frac{4}{R_{m}}
\int_{x_{_3}^{v}}^{x_{_3}}H_{v}\left( x_{_3}\right) \, h_{_3} dx_{_3}+ 2\,\int_{x_{_3}^{v}}^{x_{_3}}\rho \left( \int_{\rho _{v}}^{\rho }\frac{%
P\left( \theta ,u\right) -P_{v}}{u^{2}}du\right) \, h_{_3}  dx_{_3}.
\end{equation*}%
For\, $x_{3}\in \left[ x_{_3}^{v},x_{_3}^{i}\right] $, the quantity $\left( 4/R_{m  }\right) \int_{x_{_3}^{v}}^{x_{_3}}H_{v}\left(x_{_3}\right) \, h_{_3}  dx_{_3}$
is negligible with respect to $H_{v}\left(x_{_3}\right) $. Taking
account of Rel. (\ref{23}), we get:
\begin{equation*}
\text{For }x_{_3}\in \left[ x_{_3}^{v},x_{_3}^{i}\right] ,\ \ \ H_{v}\left(
x_{_3}\right) =\sqrt{2\,\lambda}\int_{\rho _{v}}^{\rho }\sqrt{u\int_{\rho _{v}}^{u}%
\frac{P\left( \theta ,u\right) -P_{v}}{y^{2}}\,dy}\   du ,
\end{equation*}%
where $\rho $ denotes the density associated with $x_{_3}$. All the same,
\begin{equation*}
\text{For }x_{_3}\in \left[ x_{_3}^{i},x_{_3}^{l}\right] ,\ \ \ H_{l}\left(
x_{_3}\right) =\sqrt{2\,\lambda}\int_{\rho _{l}}^{\rho }\sqrt{u\int_{\rho _{l}}^{u}%
\frac{P\left( \theta ,u\right) -P_{l}}{y^{2}}\,dy} \ du.
\end{equation*}%
In the capillary layer, the pressure $P$ is a function of $\theta $
depending on the coordinates $x_{_1}$\ and $x_{_2};\, {\rm{grad}_{tg}}\theta $ is
bounded as ${\rm{grad}_{tg}}R_{m}$ and ${\rm{grad}_{tg}}H_{j}\left(
x_{_3}\right) $ where $j\in \left\{ v,l\right\} $. In the liquid and vapour
bulks, ${\rm{grad}_{tg}}P_{l}$ and ${\rm{grad}_{tg}}P_{v}$ are bounded. By
taking   account of Eqs. (\ref{A12},  \ref{A19}, \ref{A22}) in
Appendices 1 and 2, and for the limit analysis of thin interfaces, Eq. (\ref{30})
yields
\begin{equation*}
\frac{\lambda}{2}\int_{x_{_3}^{v}}^{x_{_3}^{l}}{\rm{grad}_{tg}}\left( \rm{grad}%
\rho \right) ^{2}h_{_3}  dx_{_3}+2\,\left[ \mu \, \textbf{D}\ \mathbf{e}_{_3}\right]
_{x_{_3}^{v}}^{x_{_3}^{l}}=0.
\end{equation*}%
But,
\begin{eqnarray*}
\frac{\lambda}{2}\int_{x_{_3}^{v}}^{x_{_3}^{l}}{\rm{grad}_{tg}}\left( \frac{1}{h_{_3}}%
\frac{\partial \rho }{\partial x_{_3}}\right) ^{2}h_{_3}  dx_{_3}
&=&\lambda\int_{\rho _{v}}^{\rho _{l}}{\rm{grad}_{tg}}\left( \frac{1}{h_{_3}}\frac{%
\partial \rho }{\partial x_{_3}}\right) \,  d\rho  \\
&=&{\rm{grad}_{tg}}\left( \lambda\int_{\rho _{v}}^{\rho _{l}}\frac{1}{h_{_3}}\frac{%
\partial \rho }{\partial x_{_3}}\,d\rho \right)  \\
&=&{\rm{grad}_{tg}}\left( \lambda\int_{\rho _{v}}^{\rho _{l}}\left( \rm{grad}%
\rho \right) ^{2}h_{_3}  dx_{_3}\right).
\end{eqnarray*}%
{Then,}\qquad\qquad\qquad\qquad\qquad
$ {\rm{grad}_{tg}}H+2\left[ \mu \, \textbf{D}\, \mathbf{e}_{3}\right] _{v}^{l}=0.
$ \\

\noindent If we additively assume  that the viscous stresses are negligible in the
vapour bulk, we get
\begin{equation}
{\rm{grad}_{tg}}H + 2\,\mu_{l}\ \textbf{D}^{l}\, \mathbf{e}_{_3}=0  \label{32}.
\end{equation}%
In
the case when we consider the interface as a surface of discontinuity, the Marangoni condition is generally  presented in the form of Eq. (\ref{32}). The
calculation is obtained without any approximation and with coefficients of
viscosity non-constant across the capillary layer.
\newline
Let us note that we have obtained the interfacial energy by using a second
gradient theory but without isothermal motions. That is the case when  strong flows  cross the capillary layer corresponding to  important phase
transitions. Moreover, the capillary layer is mobile and this fact answers to the
Birkoff question   \cite{Birkhoff}.
\section{Concluding remark}
Equation (\ref{14}) gives the value of the viscous dynamical surface tension. We assume that $H$-value in dynamics is closely the same that at equilibrium. If we consider the case of water at $20^\circ$ Celsius, in c.g.s. units, $\nu_l= \mu_l/\rho_l =0.01$ and $\nu_v= \mu_v/\rho_v =0.15$. In the case of Stokes's hypothesis, $\eta =-(2/3)\mu$ and $K-H= -0.093 \times Q$.\\ For $u_l = 1\, cm/s$ corresponding to a very strong mass flow, the difference between $K$ and $H$ is not  observable  far from the critical point.
\newline
\newline

{\footnotesize
\section{Appendix 1: Orthogonal line coordinates}

\subsection{ Preliminaries  \cite
{Germain1,Aris}}

The effective thickness of a liquid-vapour interface is of nanometer range;
the other dimensions are microscopic at least. The surfaces of  equal mass density
modeling the interfacial layer can be considered as parallel
surfaces. In the
interfacial layer, the surface of equal density and the normal lines are
together a  triple orthogonal system and the intersection of the associated surfaces of the system
are the lines of curvature.
The notations are the following: scalars $x_{_1},x_{_2},x_{_3}$ denote the
curvilinear coordinates;  $\mathbf{x}\equiv\left(
x_{_1},x_{_2},x_{_3}\right) ^{T}$, where superscript $^{T}$ denotes the
transposition.  At each point $\textbf{M}$ of the interface, vectors $\mathbf{e}_{_1},\mathbf{e}_{_2},\mathbf{e}_{_3}$
denote the direct orthonormal vectors which are tangent to the coordinate lines. Vector $\mathbf{e}_{_3}$ represents the unit
normal vector collinear to $\rm{grad}\ \rho $ and directed along the increasing density. The elementary
displacement of  point $\textbf{M}$ is such that%
\begin{equation*}
d\textbf{M}=h_{_1}dx_{_1}\,\mathbf{e}_{_1}+h_{_2}dx_{_2}\,\mathbf{e}_{_2}+h_{_3}dx_{_3}\,%
\mathbf{e}_{_3}.
\end{equation*}%
We deduce in classical  derivative notations
\begin{equation*}
\frac{\partial \mathbf{e}_{_1}}{\partial x_{_1}}=-\frac {h_{_{1,2}}}{h_{_2}}\,%
\mathbf{e}_{_2}-\frac{h_{_{1,3}}}{h_{_3}}\,\mathbf{e}_{_3}.
\end{equation*}%
 We denote $ds_{_i}=h_{_i}dx_{_i},\ \ i  \in \{1,2,3\}$,
\begin{equation*}
\frac{\partial \mathbf{e}_{_1}}{\partial s_{_1}}=r_{_{1,2}}\mathbf{e}%
_{_2}+r_{_{1,3}}\mathbf{e}_{_3}\mathrm{\ \ with\ \ }r_{_{1,2}}=-\frac{%
h_{_{1,2}}}{h_{_1}h_{_2}}\mathrm{\ \ and\ \ }r_{_{1,3}}=-\frac{h_{_{1,3}}}{%
h_{_1}h_{_3}} . \label{A1}
\end{equation*}
For surfaces $\Sigma _{_{1,2}}$ generated by the two first coordinate lines, $%
r_{_{1,2}}$ and $r_{_{1,3}}$ are respectively the geodesic curvature and the
normal curvature of the first coordinate line.  Moreover,
\begin{equation*}
\frac{\partial \mathbf{e}_{_2}}{\partial s_{_1}}=-r_{_{1,2}}\mathbf{e}_{_1}%
\mathrm{\ \ and\ \ }\frac{\partial \mathbf{e}_{_3}}{\partial s_{_1}}%
=-r_{_{1,3}}\mathbf{e}_{_2} . \label{A2}
\end{equation*}
In the same way, for $i\neq j$ and belonging to $\left\{1,2,3\right\} $, we
denote \
$\displaystyle
r_{_{i,j}}=-{h_{_{i,j}}}/({h_{_i}h_{_j}}).
$
We get  analog relations   for the partial derivatives with respect to
the two last coordinates. Let us note that when surfaces $\Sigma _{_{1,2}}$ are
parallel surfaces, then $\partial \mathbf{e}_{_3}/\partial s_{_3}=0$ and
consequently \cite{Kobayashi},
\begin{equation}
r_{_{3,1}}=r_{_{3,2}}=0 \,;  \label{A3}
\end{equation}
 moreover, $\partial \mathbf{e}_{_1}/\partial s_{_3}=\partial \mathbf{e}%
_{_2}/\partial s_{_3}=0$.
\newline
In the interfacial layer, $\mathbf{e}_{_1},\mathbf{e}_{_2},\mathbf{e}_{_3}$ are
uniquely function of $x_{_1},x_{_2}$. Vectors $\mathbf{e}_{_1},\mathbf{e}_{_2}$
are the directions of the curvature lines of the surfaces of equal density.
For $i\neq j$ and belonging to $\left\{1,2,3\right\} $, $r_{_{i,j}}$ are
continuous functions of coordinates $x_{_1},x_{_2},x_{_3}$. For the limit
analysis of thin interfaces, $\ h_{_1},\ h_{_2},\mathbf{e}_{_1},\mathbf{e}_{_2},%
\mathbf{e}_{_3},r_{_{i,j}}$ can be assumed to be constant across the interfacial layer
and along the coordinate line $x_{_3}$.

\subsection{ Calculus of $\Delta  \rho $        }

For all vector fields $\mathbf{v}$ and $\mathbf{w}$,
\begin{equation*}
\text{rot}\left( \mathbf{v\times w}\right) =\mathbf{v}\,\rm{div}\mathbf{w-w%
}\,\rm{div}\mathbf{v+}\frac{\partial \mathbf{v}}{\partial \mathbf{x}}\mathbf{%
w-}\frac{\partial \mathbf{w}}{\partial \mathbf{x}}\mathbf{v},
\end{equation*}%
where $\partial /\partial \mathbf{x}$ is the gradient operator. Let us choose  $%
\mathbf{v}=\mathbf{e}_{_3}$ and $\mathbf{w}= \rm{grad}\,\rho $\,;\, then,%
\begin{equation*}
\text{rot}\left( \mathbf{e}_{_3}\mathbf{\times }\rm{grad}\rho \right) =
\mathbf{e}_{_3}\,\Delta \rho  - \rm{grad}\,\rho\  \rm{div}\,\mathbf{e}_{_3}%
 + \frac{\partial \mathbf{e}_{_3}}{\partial \mathbf{x}}\rm{grad}\,\rho
 -\frac{\partial \left( \rm{grad}\,\rho \right) }{\partial \mathbf{x}%
}\,\mathbf{e}_{_3},
\end{equation*}%
$\rm{div}\,\mathbf{e}_{_3}= -2/R_{m}$ where $R_{m}$ is
the mean curvature radius of surfaces $\Sigma _{_{1,2}}$ following the  direction $%
\mathbf{e}_{_3}$, orthonormal vector $\mathbf{e}_{_3}$ is collinear to $\rm{grad}\,\rho$   and \   $\mathbf{e}_{_3}^{T}\,\partial \mathbf{e}%
_{_3}/\partial \mathbf{x}=\mathbf{0}$; consequently we get
\begin{equation*}
\Delta \rho =-\frac{2}{R_{m}}\,\mathbf{e}_{_3}^{T}\,\rm{grad}\rho +\mathbf{e}%
_{_3}^{T}\,\frac{\partial \left( \rm{grad}\rho \right) }{\partial \mathbf{x}}%
\mathbf{e}_{_3},
\end{equation*}%
and finally
\begin{equation}
\Delta \rho =-\frac{2}{R_{m}}\frac{1}{h_{_3}}\frac{\partial \rho }{\partial
x_{_3}}+\frac{1}{h_{_3}}\left( \frac{1}{h_{_3}}\frac{\partial \rho }{\partial
x_{_3}}\right)_{,3}  \label{A4}.
\end{equation}
For the limit analysis of thin interfaces, $R_{m}$ is constant
across the interfacial layer or along the coordinate line $x_{_3}$.

\subsection{Representation of the deformation velocity tensor}

\bigskip Scalars $u_{_1},\ u_{_2},\ u_{_3}$ denote the components of the fluid
velocity $\textbf{u}$ in the system $\left( x_{_1},x_{_2},x_{_3}\right) $. The components of
the deformation velocity tensor are
\begin{eqnarray}
D_{_{11}} &=&\frac{u_{_{1,1}}}{h_{_1}}-r_{_{1,2}}u_{_2}-r_{_{1,3}}u_{_3} , \notag \\
D_{_{12}} &=&\frac{1}{2}\left( \frac{u_{_{1,2}}}{h_{_2}}+\frac{u_{_{2,1}}}{h_{_1}}%
+r_{_{1,2}}u_{_1}+r_{_{2,1}}u_{_2}\right),  \label{A5} \\
D_{_{13}} &=&\frac{1}{2}\left( \frac{u_{_{1,3}}}{h_{_3}}+\frac{u_{_{3,1}}}{h_{_1}}%
+r_{_{1,3}}u_{_1}+r_{_{3,1}}u_{_3}\right) .  \notag
\end{eqnarray}
There exist six other expressions obtained by circular permutation of indices
1,2,3. From Eq. (\ref{A3}), we get
\begin{equation}
D_{_{33}}=\frac{u_{_{3,3}}}{h_{_3}} . \label{A6}
\end{equation}%

\subsection{Kinematics of interfaces}
We denote
$u = u_{_3}$
 the fluid velocity with respect to a surface of
iso-density and   by $Q$ the mass flow through the interface,
\begin{equation}
Q =\rho\, u   \label{A8}.
\end{equation}%
In the capillary layer   $Q$ is only a function on $x_{_1},x_{_2}$.
Let us note that $Q=0$ is equivalent to $\rm{div}\,\mathbf{u}=0$. From
relations (\ref{A6}) and (\ref{A8}), we get,%
\begin{eqnarray}
\left[ \eta\, u_{_3}\right] _{x_{_3}^{v}}^{x_{_3}} &=& Q\left[ \frac{\eta }{%
\rho }\right] _{x_{_3}^{v}}^{x_{_3}} , \label{A9} \\
\left[ \left( \eta +2\,\mu \right) D_{_{33}}\right] _{x_{_3}^{v}}^{x_{_3}} &=&%
\left[ -\frac{1}{h_{_3}}\frac{Q}{\rho ^{2}}\left( \eta +2\,\mu \right) \frac{%
\partial \rho }{\partial x_{_3}}\right] _{x_{_3}^{v}}^{x_{_3}}.  \label{A10}
\end{eqnarray}%
When $Q=0$, then $\rm{div}\,\mathbf{u}=0$  and $D_{_{33}}=0$.\ Consequently,
\begin{equation}
\left[ \mu\, \textbf{D}\,\mathbf{e}_{_3}\right] _{x_{_3}^{v}}^{x_{_3}^{l}}=\left[ \mu D_{_{13}}
\mathbf{e}_{_1}+\mu\, D_{_{23}} \mathbf{e}_{_2}\right] _{x_{_3}^{v}}^{x_{_3}^{l}}.
\label{A12}
\end{equation}

\section{Appendix 2: Conditions associated with the dissipative
function}

\emph{In the interfacial layer, we also assume} that $u_{_1},u_{_2},u_{_3}$ and $\eta ,\
\mu $ are bounded and have partial derivative bounded with respect to $%
x_{_1},x_{_2}$.

\subsection{Property 1:}

\emph{The tangential components of the fluid velocity are continuous through
the interface.
Consequently for the limit analysis of thin interfaces, we use the approximation that $u_{_1}$ and $u_{_2}$
are constant through the interface.}
\newline
\newline
The proof of this property comes from the dissipative function $\psi $
associated with the viscous stress tensor.
\begin{equation*}
\psi =\frac{1}{2}\left(\eta\, \left( \text{tr}\,\textbf{D}\right)^{2}+ 2\,\mu\, \text{tr}\left(
 \textbf{D}^{2}\right)\right).  \label{A13}
\end{equation*}%
With the notations of Appendix 1,
\begin{equation*}
\psi = \frac{\left( \eta +2\,\mu \right) }{2}\left(
D_{_{11}}^{2}+D_{_{22}}^{2}+D_{_{33}}^{2}\right) +2\,\mu \left(
D_{_{12}}^{2}+D_{_{13}}^{2}+D_{_{23}}^{2}\right)
+
\eta \left\{ D_{_{11}}\,D_{_{22}}+D_{_{33}}\left( D_{_{11}}+D_{_{22}}\right) \right\}.
\end{equation*}%
The dissipative function must have an integral $\int_{x_{_3}^{v}}^{x_{_3}^{l}}\left( \eta \left( \text{tr}\,
\textbf{D}\right) ^{2}+ 2\,\mu\, \text{tr}\left( \, \textbf{D}  ^{2}\right)\right) h_{_3}  dx_{_3}$ bounded in the layer.   With the hypothesis of Section 7.3   on the partial derivatives of component velocity, we deduce that $%
D_{_{11}}, D_{_{12}},  D_{_{22}}$ are bounded across the interface and due to (Eq. (\ref{A6})),
\begin{equation*}
\int_{x_{_3}^{v}}^{x_{_3}^{l}}\eta\, D_{_{33}}\left( D_{_{11}}+D_{_{22}}\right) h_{_3}
dx_{_3}=\int_{v_{_3}^{v}}^{v_{_3}^{l}}\eta \left( D_{_{11}}+D_{_{22}}\right)
du_{_3}
\end{equation*}%
is bounded.  Moreover, Eq. (\ref{A3}) implies
\begin{equation*}
D_{_{13}}=\frac{1}{2}\left( \frac{u_{_{1,3}}}{h_{_3}}+\frac{u_{_{3,1}}}{h_{_1}}%
+r_{_{1,3}}u_{_1}\right).
\end{equation*}%
Due to $\mu >0$ and $\eta +2\,\mu >0$\  \cite{Landau}, and the fact   that in
the integral of $D_{_{13}}^{2}$, the term
\begin{equation*}
\int_{x_{_3}^{v}}^{x_{_3}^{l}}\mu \frac{u_{_{1,3}}}{h_{_3}}\left( \frac{u_{_{3,1}}}{%
h_{_1}}+r_{_{1,3}}u_{_1}\right) h_{_3}  dx_{_3}=\int_{u_{_1}^{v}}^{u_{1}^{l}}\mu
\left( \frac{u_{_{3,1}}}{h_{_1}}+r_{_{1,3}}u_{_1}\right)   du_{_1}
\end{equation*}%
is bounded, the term $\int_{x_{_3}^{v}}^{x_{_3}^{l}}\mu \frac{u_{_{1,3}}^{2}}{%
h_{_3}}h_{_3}  dx_{_3}$ must be bounded. But $\int_{x_{_3}^{v}}^{x_{_3}^{l}}\frac{%
u_{_{1,3}}^{2}}{h_{_3}}h_{_3}   dx_{_3}$ is minimum when $u_{_{1,3}}$ is
independent of $x_{_3}$, that is to say, $ u_{_{1,3}} = \frac {
u_{_1}^{l}-u_{_1}^{v}} {h}$, where $h$ is the interfacial thickness. Then,%
\begin{equation*}
\int_{x_{_3}^{v}}^{x_{_3}^{l}}\frac{u_{_{1,3}}^{2}}{h_{_3}}h_{_3}\ dx_{_3}\geq \frac{%
\left( u_{_1}^{l}-u_{_1}^{v}\right) ^{2}}{h}.
\end{equation*}%
Additively, $\mu > 0$ implies
\begin{equation*}
\int_{x_{_3}^{v}}^{x_{_3}^{l}}2\,\mu \frac{u_{_{1,3}}^{2}}{h_{_3}}h_{_3}\ dx_{_3}\geq
2\,\mu _{\min }\frac{\left( u_{_1}^{l}-u_{_1}^{v}\right) ^{2}}{h}.
\end{equation*}%
Consequently, for $u_{_1}^{l}\neq u_{_1}^{v}$, the dissipative function goes
to infinity when $h$ goes to zero. The component $u_{_1}$ of the velocity is
continuous across the interface and it is the same for component $u_{_2}$.

\subsection{Property 2:}

\emph{For a motion normal to the interface,}
\begin{equation}
\int_{x_{_3}^{v}}^{x_{_{_3}}}\left( \frac{\partial \left( \eta\, \rm{div}%
\mathbf{u}\right) }{\partial x_{_{3}}}+2\,h_{_3}\left( \rm{div}\left( \mu
\,\textbf{D}\right) \right) _{3}\right) dx_{_3}=\left[ \left( \eta +2\,\mu \right)
D_{_{33}}\right] _{x_{_3}^{v}}^{x_{_3}}-\frac{2}{R_{m}}\left[ \eta\, u_{_3}\right]
_{x_{_3}^{v}}^{x_{_3}}.  \label{A14}
\end{equation}
\newline

\noindent We first get,
$\displaystyle
  \ \int_{x_{_3}^{v}}^{x_{_{_3}}}\frac{\partial \left( \eta\, \rm{div}\,\mathbf{u}%
\right) }{\partial x_{_{3}}}dx_{_3}=\left[ \eta\, \rm{div}\,\mathbf{u}\right]
_{x_{_3}^{v}}^{x_{_3}}.$\\
Moreover, due to Property 1, $\displaystyle\frac{u_{_{1,1}}}{h_{_1}},%
\frac{u_{_{2,1}}}{h_{_1}},\frac{u_{_{1,2}}}{h_{_2}}$ and $\displaystyle\frac{u_{_{2,2}}}{%
h_{2}}$ are null across the interface. From Rel. (\ref{A5}),
\begin{eqnarray*}
\left[ \eta\, D_{_{11}}\right] _{x_{_3}^{v}}^{x_{_3}} &=&\frac{u_{_{1,1}}}{h_{_1}}%
\left[ \,\eta\, \right] _{x_{_3}^{v}}^{x_{_3}}-r_{_{1,2}}u_{_2}\left[\, \eta\,
\right] _{x_{_3}^{v}}^{x_{_3}}-r_{_{1,3}}\left[ \eta\, u_{_3}\right]
_{x_{_3}^{v}}^{x_{_3}}, \\
\left[ \eta\, D_{_{22}}\right] _{x_{_3}^{v}}^{x_{_3}} &=&\frac{u_{_{2,2}}}{h_{_2}}%
\left[ \,\eta\, \right] _{x_{_3}^{v}}^{x_{_3}}-r_{_{2,1}}u_{_1}\left[ \,\eta \,
\right] _{x_{_3}^{v}}^{x_{_3}}-r_{_{2,3}}\left[ \eta\, u_{_3}\right]
_{x_{_3}^{v}}^{x_{_3}}
\end{eqnarray*}%
and consequently%
\begin{equation*}
\left[ \eta\, \rm{div}\mathbf{u}\right] _{x_{_3}^{v}}^{x_{_3}}=-\frac{2}{%
R_{m}}\left[ \eta \,u_{_3}\right] _{x_{_3}^{v}}^{x_{_3}}+\left[\, \eta\, \right]
_{x_{_3}^{v}}^{x_{_3}}\left( \frac{u_{_{1,1}}}{h_{_1}}+\frac{u_{_{2,2}}}{h_{_2}}%
-r_{_{1,2}}u_{_2}-r_{_{2,1}}u_{_1}\right) +\left[ \eta D_{_{33}}\right]
_{x_{_3}^{v}}^{x_{_3}}.
\end{equation*}%
When $u_{_1}=u_{_2}=0$,
\begin{equation*}
\left[ \eta\, \rm{div}\,\mathbf{u}\right] _{x_{_3}^{v}}^{x_{_3}}=\left[
\eta\, D_{_{33}}-\frac{2}{R_{m}}\eta\, u_{_3}\right] _{x_{_3}^{v}}^{x_{_3}}.
\label{A16}
\end{equation*}%
Moreover,%
\begin{eqnarray*}
\left( \rm{div}\,\mu\, \textbf{D}\right) _{3} &=&\frac{1}{h_{_1}h_{_2}h_{_3}}\left( \left(
h_{_1}h_{_2}\mu D_{_{33}}\right) _{,3}+\left( h_{_2}h_{_3}\mu D_{_{31}}\right)
_{,1}+\left( h_{_3}h_{_1}\mu D_{_{32}}\right) _{,2}\right)  \label{A17} \\
&&+\mu D_{_{13}}\frac{h_{_{3,1}}}{h_{_3}h_{_1}}+\mu D_{_{23}}\frac{h_{_{3,2}}}{h_{_3}h_{_2}}%
-\mu D_{_{11}}\frac{h_{_{1,3}}}{h_{_1}h_{_3}}-\mu D_{_{22}}\frac{h_{_{2,3}}}{h_{_2}h_{_3}} .
\end{eqnarray*}%
Taking   account of Eq. (\ref{A3}), we obtain%
\begin{eqnarray*}
\left( \rm{div}\,\mu\, \textbf{D}\right) _{3} &=&\frac{1}{h_{_1}h_{_2}h_{_3}}\left( \left(
h_{_1}h_{_2}\,\mu\, D_{_{33}}\right) _{,3}+\left\{ \frac{h_{_2}h_{_3}}{2}\left( \frac{%
\mu\, u_{_{1,3}}}{h_{_3}}+\frac{\mu\, u_{_{3,1}}}{h_{_1}}+\mu\, u_{_1}r_{_{1,3}}+\mu\,
u_{_3}r_{_{3,1}}\right) \right\} _{,1}\right. \\
&&+\left. \left\{ \frac{h_{_3}h_{_1}}{2}\left( \frac{\mu\, u_{_{3,2}}}{h_{_2}}+\frac{%
\mu\, u_{_{2,3}}}{h_{_3}}+\mu\, u_{_2}r_{_{2,3}}+\mu\, u_{_3}r_{_{3,2}}\right) \right\}
_{,2}\right) \\
&&+r_{_{1,3}}\left( \frac{\mu\, u_{_{1,1}}}{h_{_1}}-\mu\, u_{_2}r_{_{1,2}}-\mu\,
u_{_3}r_{_{1,3}}\right) +r_{_{2,3}}\left( \frac{\mu\, u_{_{2,2}}}{h_{_2}}-\mu\,
u_{_3}r_{_{2,3}}-\mu\, u_{_1}r_{_{2,1}}\right) .
\end{eqnarray*}%
The only non-bounded term across the interface is
\begin{equation*}
\frac{1}{h_{_ 1}h_{_2}h_{_3}}\left( \left( h_{_1}h_{_2}\mu D_{_{33}}\right)
_{,3}+\left\{ \frac{h_{_2}h_{_3}}{2}\frac{\mu\, u_{_{1,3}}}{h_{_3}}\right\}
_{,1}+\left\{ \frac{h_{_3}h_{_1}}{2}\frac{\mu\, u_{_{2,3}}}{h_{_3}}\right\}
_{,2}\right) .
\end{equation*}%
Consequently,%
\begin{eqnarray*}
\int_{x_{_3}^{v}}^{x_{_{3}}}\left( \rm{div}\,\mu\, \textbf{D}\right) _{3}h_{_3}\ dx_{_3}
&=&\int_{x_{_3}^{v}}^{x_{_{3}}}\frac{1}{h_{_1}h_{_2}}\left\{ \left(
h_{_1}h_{_2}\mu\, D_{_{33}}\right) _{,3}+\left( \frac{h_{_2}}{2}\mu\, u_{_{1,3}}\right)
_{,1}+\left( \frac{h_{_1}}{2}\mu\, u_{_{2,3}}\right) _{,2}\right\}  dx_{_3} \\
&=&\left[ \mu\, D_{_{33}}\right] _{x_{_3}^{v}}^{x_{_3}}+\int_{x_{_3}^{v}}^{x_{_{3}}}%
\frac{1}{h_{_1}h_{_2}}\left\{ \left( \frac{h_{_2}}{2}\mu\, u_{_{1,3}}\right)
_{,1}+\left( \frac{h_{_1}}{2}\mu\, u_{_{2,3}}\right) _{,2}\right\}  dx_{_3}.
\end{eqnarray*}%
When  no tangential motion appears along the interface, $
u_{_1}=u_{_2}=0$ and $%
u_{_{1,3}}=u_{_{2,3}}=0 $. Consequently, $ \int_{x_{_3}^{v}}^{x_{_{3}}}
\left( \rm{div}\,\mu\, \textbf{D}\right) _{3}h_{_3}\ dx_{_3}=%
\left[ \mu\, D_{_{33}}\right] _{x_{_3}^{v}}^{x_{_3}}
$ and
we get Rel. (\ref{A14}).

\subsection{Property 3:} \emph{Across the interface,}
\begin{equation}
\int_{x_{_3}^{v}}^{x_{_{3}}^{l}}\left( \rm{div}\,\mu\, \textbf{D}\right) _{tg}h_{_3}
dx_{_3}=\left[ \mu\, D_{_{13}}\,\mathbf{e}_{_1}+\mu\, D_{_{23}}\,\mathbf{e}_{_2}\right]
_{x_{_3}^{v}}^{x_{_{3}}^{l}}.  \label{A19}
\end{equation}
\newline

\noindent This relation comes from the following calculations
\begin{eqnarray*}
\left( \rm{div}\,\mu\, \textbf{D}\right) _{1} &=&\frac{1}{h_{_1}h_{_2}h_{_3}}\left( \left(
h_{_2}h_{_3}\mu\, D_{_{11}}\right) _{,1}+\left( h_{_3}h_{_1}\mu\, D_{_{12}}\right)
_{,2}+\left( h_{_1}h_{_2}\mu\, D_{_{13}}\right) _{,3}\right) \\
&&+\mu\, D_{_{21}}\frac{h_{_{1,2}}}{h_{_1}h_{_2}}+\mu\, D_{_{31}}\frac{h_{_{1,3}}}{h_{_1}h_{_3}}%
-\mu\, D_{_{22}}\frac{h_{_{2,1}}}{h_{_1}h_{_2}}-\mu\, D_{_{33}}\frac{h_{_{3,1}}}{h_{_1}h_{_3}}
\end{eqnarray*}
\begin{eqnarray*}
&=&\frac{1}{h_{_1}h_{_2}h_{_3}}\left( \left\{ \mu\, h_{_2}h_{_3}\left( \frac{u_{_{1,1}}%
}{h_{_1}}-r_{_{1,2}}u_{_2}-r_{_{1,3}}u_{_3}\right) \right\} _{,1}\right. \\
&&+\left. \left\{ \frac{\mu\, h_{_3}h_{_1}}{2}\left( \frac{u_{_{1,2}}}{h_{_2}}+\frac{%
U_{_{2,1}}}{h_{_1}}+r_{_{1,2}}u_{_1}+r_{_{2,1}}u_{_2}\right) \right\} _{,2}+\left\{ \mu\,
h_{_1}h_{_2}D_{_{13}}\right\} _{,3}\right) \\
&&-\,\frac{\mu }{2}\left( \frac{u_{_{1,2}}}{h_{_2}}+\frac{u_{_{2,1}}}{h_{_1}}%
+r_{_{1,2}}u_{_1}+r_{_{2,1}}u_{_2}\right) r_{_{1,2}}
 -\frac{\mu }{2}\left( \frac{u_{_{1,3}}%
}{h_{_3}}+\frac{u_{_{3,1}}}{h_{_1}}+r_{_{1,3}}u_{_1}+r_{_{3,1}}u_{_3}\right) r_{_{1,3}} \\
&& +\, \mu \left( \frac{u_{_{2,2}}}{h_{_2}}-r_{_{2,3}}u_{_3}-r_{_{2,1}}u_{_1}\right)
r_{_{2,1}}+\mu \left( \frac{u_{_{3,3}}}{h_{_3}}-r_{_{3,1}}u_{_1}-r_{_{3,2}}u_{_2}\right)
r_{_{3,1}}.
\end{eqnarray*}
Taking  account of Rel. (\ref{A3}), the non-bounded term  across the
interface is
\begin{equation*}
\frac{\left( h_{_1}h_{_2}\mu\, D_{_{13}}\right) _{,3}}{h_{_1}h_{_2}h_{_3}}-\mu \frac{%
r_{_{1,3}}}{2}\frac{u_{_{1,3}}}{h_{_3}}.
\end{equation*}
Consequently,
\begin{eqnarray*}
\int_{x_{_3}^{v}}^{x_{_{3}}^{l}}\left( \rm{div}\,\mu\, \textbf{D}\right) _{1}\ h_{_3}
dx_{_3} &=&\int_{x_{_3}^{v}}^{x_{_{3}}^{l}}\frac{1}{h_{_1}h_{_2}}\left(
h_{_1}h_{_2}\mu D_{_{13}}\right) _{,3}\ dx_{_3}-\int_{x_{_3}^{v}}^{x_{_{3}}^{l}}\mu
\frac{r_{_{1,3}}u_{_1}}{2}u_{_{1,3}}  dx_{_3} \\
&=&\ \left[ \mu\, D_{_{13}}\right] _{x_{_3}^{v}}^{x_{_{3}}^{l}}-\frac{r_{_{1,3}}}{2}%
\int_{u_{_1}^{v}}^{u_{_{1}}^{l}}\mu\, u_{_1}   du_{_1}.
\end{eqnarray*}
With the hypothesis of the limit analysis of thin interfaces, the term $\frac{r_{_{1,3}}}{2}%
\int_{u_{_1}^{v}}^{u_{_{1}}^{l}}\mu\, u_{_1}   du_{_1}$ is null and from an analogous calculation for $
\int_{x_{_3}^{v}}^{x_{_{3}}^{l}}\left( \rm{div}\,\mu\, \textbf{D}\right) _{2}\, h_{_3}\,
dx_{_3}$ we get Rel. (\ref{A19}).

\subsection{Property 4:}

\emph{For a null flow across the interface,}
\begin{equation}
\int_{x_{3}^{v}}^{x_{_{3}}^{l}}\frac{\left( \rm{div}\,\mu\, \textbf{D}\right) _{3}}{%
\rho }\ h_{_3}\, dx_{_3}=0  .\label{A21}
\end{equation}
\\

\noindent When $Q=0,$ we get $D_{_{33}}=0$ (see Eqs. (\ref{A5}) and (\ref{A6})); we deduce
\begin{equation*}
\int_{x_{_3}^{v}}^{x_{_{3}}^{l}}\frac{\left( \rm{div}\,\mu\, \textbf{D}\right) _{3}}{%
\rho}\ h_{_3}\ dx_{_3}=\int_{x_{_3}^{v}}^{x_{_{3}}^{l}}\frac{1}{\rho\, h_{_1}h_{_2}%
}\left( \left( \frac{\mu\, h_{_2}}{2}u_{_{1,3}}\right) _{,1}+\left( \frac{\mu\, h_{_1}%
}{2}u_{_{2,3}}\right) _{,2}\right) dx_{_3}.
\end{equation*}
For   the limit analysis of thin interfaces, across the interface
$u_{_{1,3}}=u_{_{2,3}}=0 $  and we get Rel. (\ref{A21}).
\subsection{Property 5:}  \emph{Across the interface,}
\begin{equation}
\int_{x_{3}^{v}}^{x_{_{3}}^{l}}\rho\, a_{tg}\, h_{_3}\, dx_{_3}=0
\label{A22}.
\end{equation}

\noindent In fact, Rel. (\ref{A3}) implies
\begin{equation*}
a_{_1}=\frac{\partial u_{_1}}{\partial t}+\frac{u_{_1}u_{_{1,1}}}{h_{_1}}+%
\frac{u_{_2}u_{_{1,2}}}{h_{_2}}+\frac{u_{_3}u_{_{1,3}}}{h_{_3}}%
-u_{_1}u_{_2}r_{_{1,2}}-u_{_1}u_{_3}r_{_{1,3}}+u_{_2}^{2}r_{_{2,1}}.
\end{equation*}
For $\partial u_{_1}/\partial t$ bounded, we immediately deduce from   the limit analysis of thin interfaces  that $%
\int_{x_{_3}^{v}}^{x_{_{3}}^{l}}\rho \, a _{_1}\, h_{_3}  dx_{_3}=0$ and the
same result for $a_{_2}$.
\subsection{Property 6:}  \emph{Through the interface,}
\begin{equation}
\int_{x_{_3}^{v}}^{x_{_{3}}}\rho\, a_{_3}\, h_{_3}\, dx_{_3}=Q^{2}\left( \frac{%
1}{\rho }-\frac{1}{\rho _{v}}\right)  \label{A23}.
\end{equation}

\noindent
The only terms non bounded across the interface are: $\partial
u_{_3}/\partial t$ and $u_{_3}u_{_{3,3}}/h_{_3}$. Consequently,%
\begin{equation*}
\int_{x_{_3}^{v}}^{x_{_{3}}}\rho \,a_{_3}\, h_{_3}\,
dx_{_3}=\int_{x_{_3}^{v}}^{x_{_{3}}}\rho \left( \frac{\partial u_{_3}}{\partial
t}+\frac{u_{_3}u_{_{3.3}}}{h_{_3}}\right) \, h_{_3}  dx_{_3}.
\end{equation*}
But $u_{_3}=Q/\rho $ and $\partial u_{_3}/\partial t=\left( \partial
Q/\partial t\right) \left( 1/\rho \right) -\left( Q^{2}/\rho ^{2}\right)
\left( \partial \rho /\partial t\right) $. If we assume that $\partial
Q/\partial t$ is bounded and
\begin{equation}
u_{_{3,3}}=-\left( Q/\rho ^{2}\right) \rho _{_{,3}},  \label{A24}
\end{equation}
then,%
\begin{equation*}
\int_{x_{_3}^{v}}^{x_{_{3}}}\rho \,a_{_3}\, h_{_3}\,
dx_{_3}=\int_{x_{_3}^{v}}^{x_{_{_3}}}\left( -\frac{Q}{\rho }\frac{\partial \rho
}{\partial t}h_{_3}-\frac{u_{_3}Q}{\rho }\rho _{_{,3}}\right) dx_{_3}.
\end{equation*}
Taking   account of Eq. (\ref{4}), we get
\begin{equation}
\frac{\partial \rho }{\partial t}+\frac{\rho }{h_{_1}h_{_2}h_{_3}}\left\{
\left( u_{_1}h_{_2}h_{_3}\right) _{,1}+\left( u_{_2}h_{_3}h_{_1}\right)
_{,2}+\left( u_{_3}h_{_1}h_{_2}\right) _{,3}\right\} +\frac{\rho _{_{,3}}}{h_{_3}}%
u_{_3}=0.  \label{A24'}
\end{equation}
From Eq. (\ref{A24}), the only non bounded term of Eq. (\ref{A24'}) is
\begin{equation*}
\frac{\partial \rho }{\partial t}+\rho \frac{u_{_{3,3}}}{h_{_3}}+\frac{\rho _{_{,3}}%
}{h_{_3}}u_{_3}\equiv\frac{\partial \rho }{\partial t} .
\end{equation*}
Its integral across the interface is null.  That is the same for the
integral of $\left( Q/\rho \right) \left( \partial \rho /\partial t\right) $.%
$\ $Consequently,%
\begin{equation*}
\int_{x_{_3}^{v}}^{x_{_{3}}}\rho \,a_{_3}\, h_{_3}\,
dx_{_3}=-\int_{x_{_3}^{v}}^{x_{_{3}}}\frac{u_{_3}Q}{\rho }\rho _{_{,3}
}\,dx_{3}=-Q^{2}\int_{x_{_3}^{v}}^{x_{_{3}}}\frac{\rho _{_{,3}}}{\rho ^{2}}\,
dx_{_3}=Q^{2}\left( \frac{1}{\rho }-\frac{1}{\rho _{v}}\right).
\end{equation*}

\subsection{Property 7:}

\emph{For a null flow across the interface,}%
\begin{equation}
\int_{x_{_3}^{v}}^{x_{_{3}}} a_{_3}\, h_{_3}  dx_{_3}=0.  \label{A25}
\end{equation}

\noindent All the same, this result is an immediate consequence   of
the limit analysis of thin interfaces.
}

\phantomsection
\vspace*{0.5cm}
{\footnotesize\begin{tabular}{rl}
\hline
& \\
$^{\star}$
&  Aix-Marseille Universit\'e, CNRS, Centrale Marseille, M2P2 UMR 7340, \\
&  13451, Marseille, France\\
& {Email}: henri.gouin@univ-amu.fr; henri.gouin@yahoo.fr\\
\end{tabular}
}

\vfill

\end{document}